\documentclass{appolb}
\usepackage{graphicx}
\usepackage{subfigure}
\usepackage{braket}

\renewcommand{\d}{\ensuremath{\mathrm{d}}}

\begin{document}
\title{Lattice Landau gauge gluon propagator at finite temperature: non-zero Matsubara frequencies and spectral densities
\thanks{Presented by P. J. Silva at Excited QCD 2017, Sintra, Portugal, 2017.}%
}

\author{Paulo J. Silva$^1$, Orlando Oliveira$^1$,\\David Dudal$^{2,3}$, Martin Roelfs$^{2}$
\address{$^1$CFisUC, Department of Physics, University of Coimbra, P--3004 516 Coimbra, Portugal \\
$^2$KU Leuven Campus Kortrijk -- KULAK, Department of Physics, Etienne Sabbelaan 53 bus 7657, 8500 Kortrijk, Belgium\\
$^3$Ghent University, Department of Physics and Astronomy, Krijgslaan 281-S9, 9000 Gent, Belgium}
}

\maketitle
\begin{abstract}
The lattice Landau gauge gluon propagator at finite temperature is computed including the non-zero Matsubara frequencies. Furthermore, the K\"all\'en-Lehmann representation is inverted and the corresponding spectral density evaluated using a Tikhonov regularisation together with the Morozov discrepancy principle. Implications for gluon confinement are discussed.
\end{abstract}
\PACS{PACS numbers come here}
  
\section{Introduction}

Recent heavy-ion experiments running e.g. at RHIC \cite{rhicexp} and CERN \cite{cernexp} motivate further theoretical studies of QCD at finite temperature and density. For pure SU(3) Yang-Mills theory at finite temperature (and zero density), lattice simulations were able to find a first-order phase transition at a critical temperature $T_c\sim 270$ MeV  \cite{Tc, nearTc}, with the gluons becoming deconfined and behaving as massive quasiparticles for temperatures above $T_c$ \cite{gluonmass}.

Here we focus on the gluon propagator, in Landau gauge, computed on the lattice, including the non-zero Matsubara frequencies. We also show preliminary results for the associated spectral density.

\section{The gluon propagator at finite temperature}

At finite temperature, the gluon propagator in Landau gauge has two form factors, the transverse (magnetic) $D_T$ and longitudinal (electric) $D_L$:
\begin{equation}
D^{ab}_{\mu\nu}(\hat{q})=\delta^{ab}\left(P^{T}_{\mu\nu} D_{T}(q_4,\vec{q})+P^{L}_{\mu\nu} D_{L}(q_4,\vec{q}) \right) 
\label{tens-struct}
\end{equation}

We aim to  compute  $D_L$ and $D_T$ for all Matsubara frequencies. This can be achieved by considering suitable combinations of $D^{aa}_{ii}(\hat{q})$ and $D^{aa}_{44}(\hat{q})$. Herein, we analyse the finite temperature lattice ensembles generated in Coimbra \cite{lca} with the help of Chroma \cite{chroma} and PFFT \cite{pfft} libraries, reported in \cite{gluonmass}. 

\begin{figure}[t] 
   \centering
   \subfigure{ \includegraphics[scale=0.23, angle=-90]{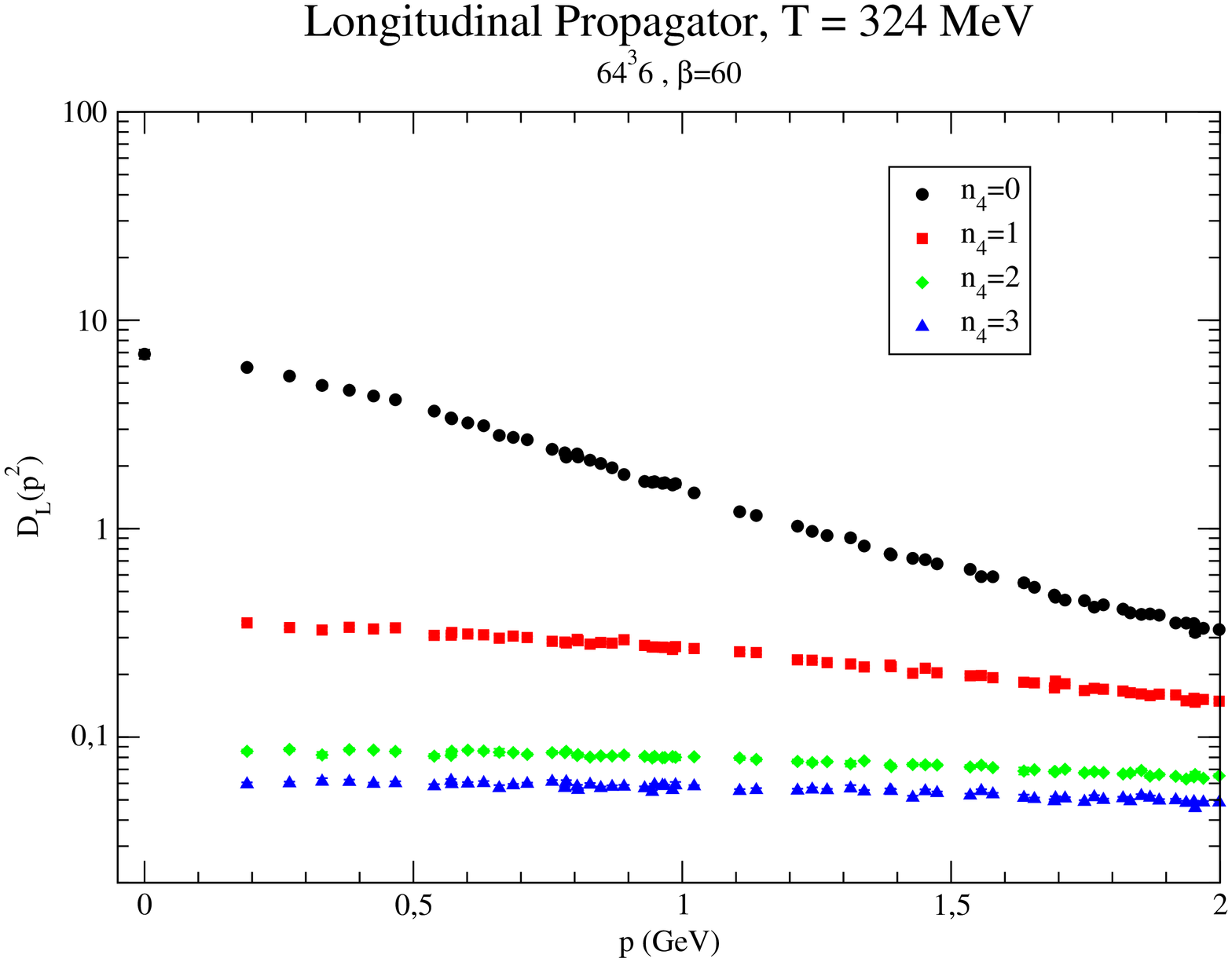} } \hspace*{-10mm}
   \subfigure{ \includegraphics[scale=0.23, angle=-90]{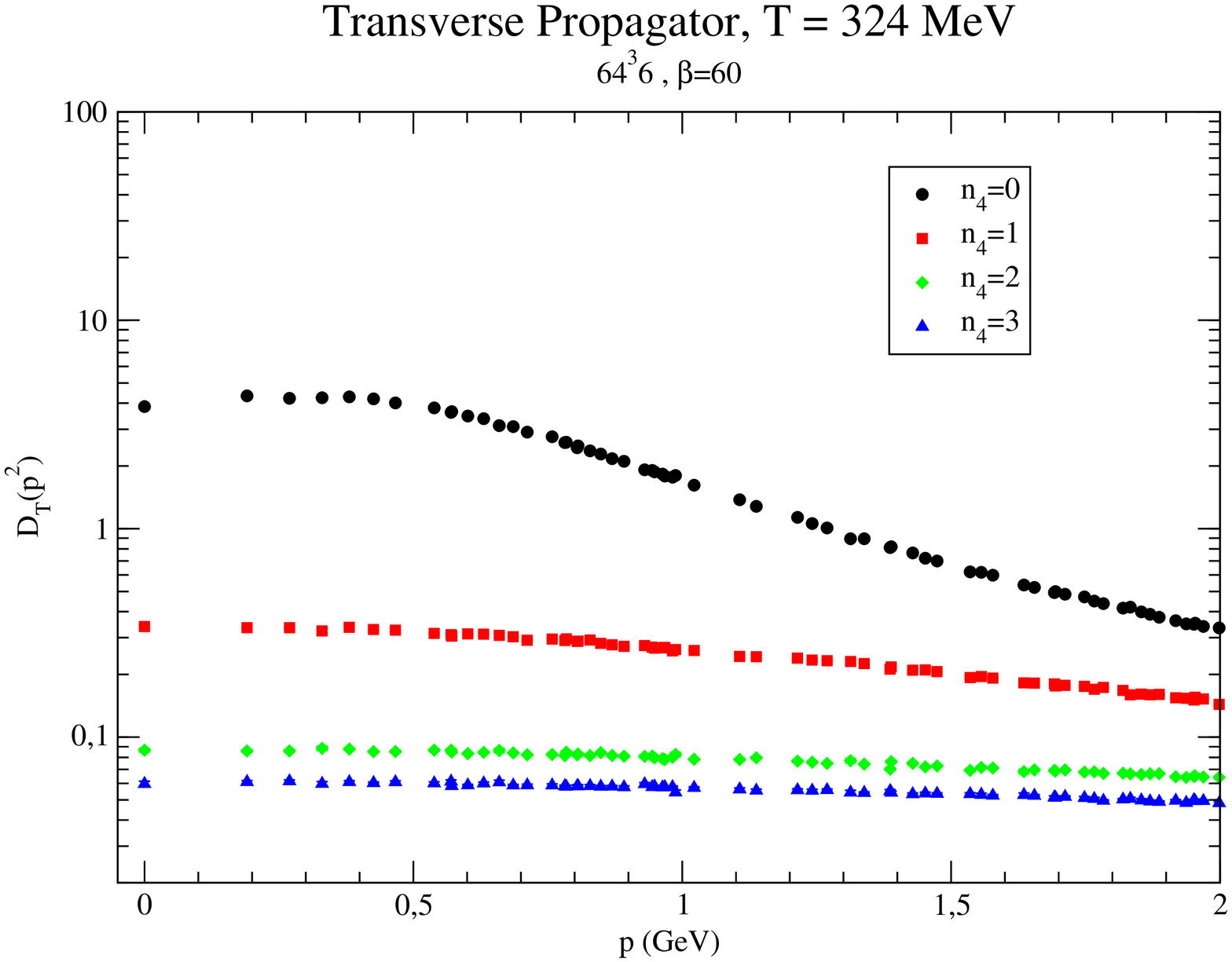} }
  \caption{Longitudinal (left) and transverse (right) components of the gluon propagator for $T=324$ MeV, including the non-zero Matsubara frequencies.}
   \label{gluonprop}
\end{figure}

Typical results for the form factors are shown in Fig. \ref{gluonprop}. For both components, we see that the propagator in the infrared region takes smaller values for higher Matsubara frequencies. This translates into larger mass scales as $q_4$ is increased.

\begin{figure}[t] 
   \centering
   \subfigure{ \includegraphics[scale=0.23, angle=-90]{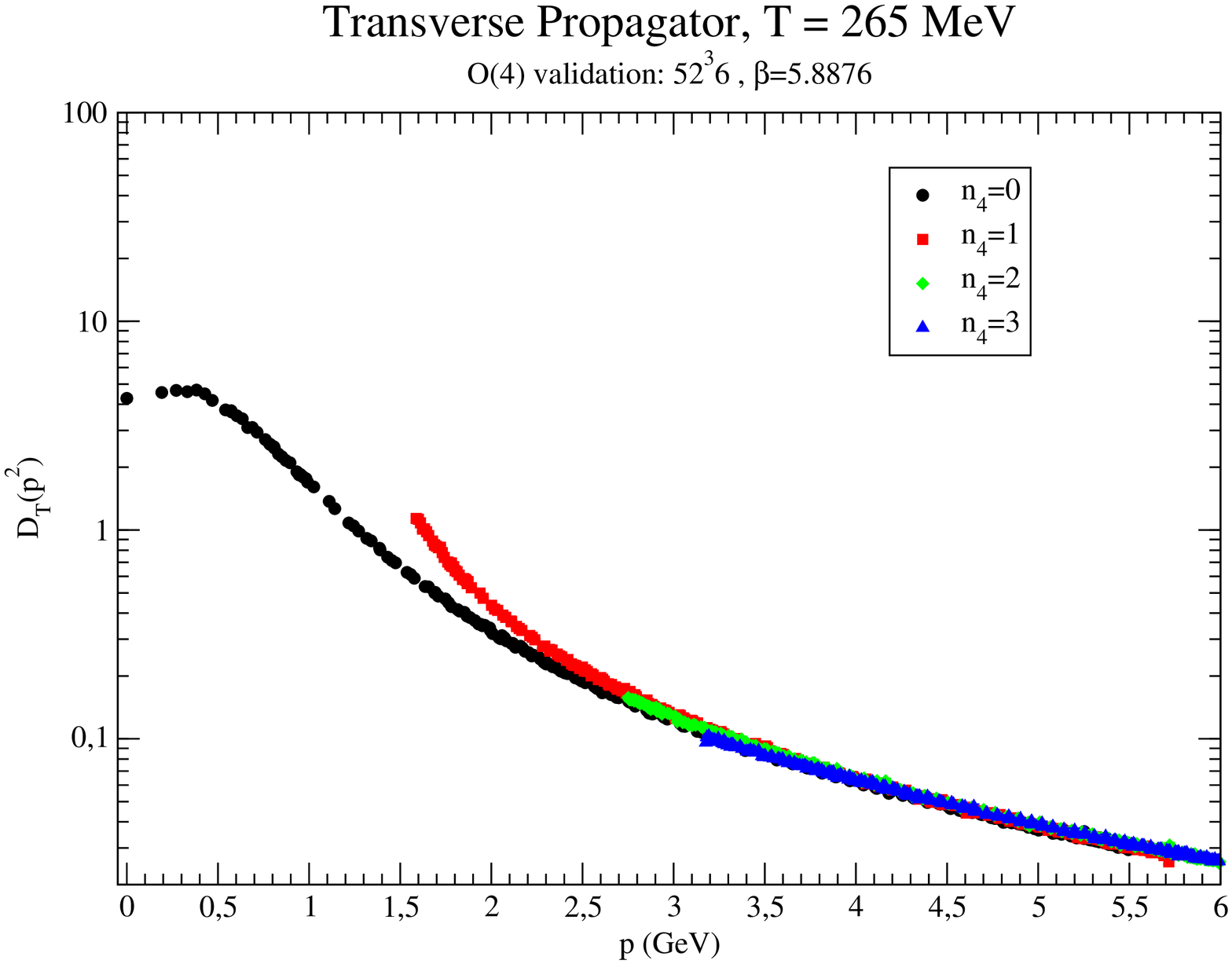} } \hspace*{-10mm}
   \subfigure{ \includegraphics[scale=0.23, angle=-90]{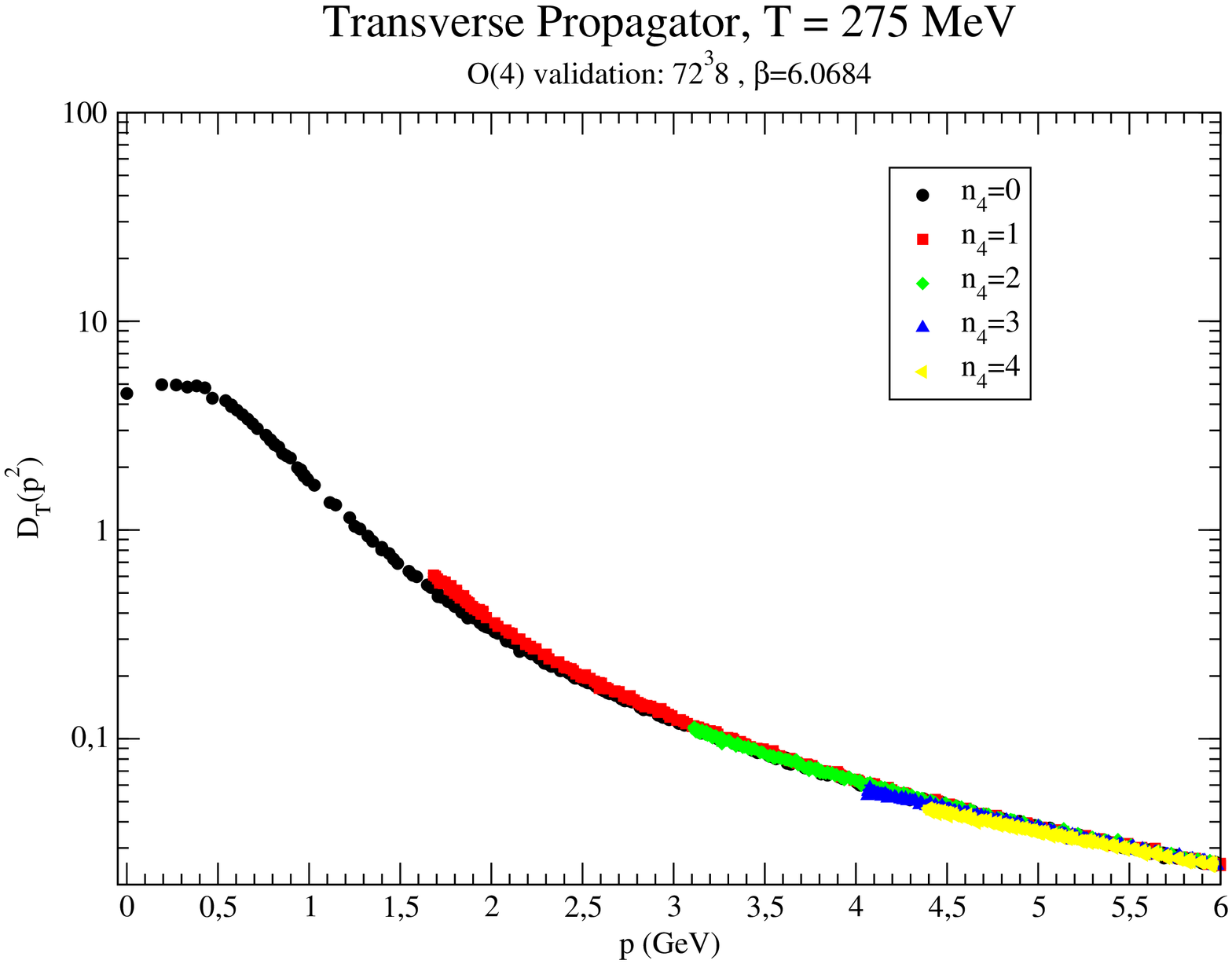} }
  \caption{$O(4)$ invariance for $D_T$ at $T=265$ MeV (left) and $T=275$ MeV (right).}
   \label{o4fig}
\end{figure}

In Fig. \ref{o4fig} we verify whether the results comply with the so-called $O(4)$ invariance, i.e. if $D(q_4,\vec{q})=D(0, q)$ with $q^2=q_4^2+\vec{q}^{\,2}$. The lattice data is compatible with the $O(4)$ invariance, except for a  small set of temperatures just below  $T_c$.

\begin{figure}[t] 
   \centering
   \subfigure{ \includegraphics[scale=0.23, angle=-90]{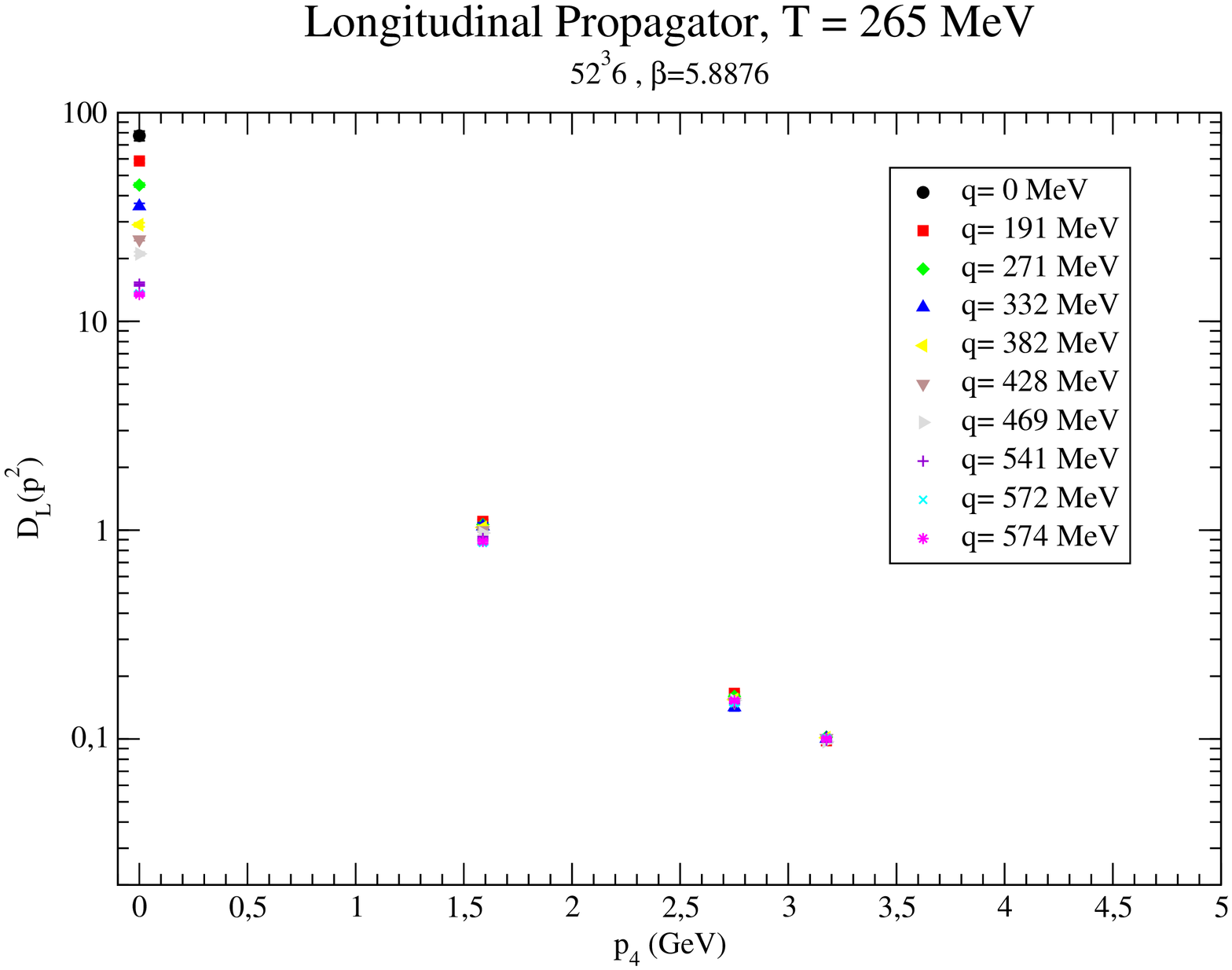} } \hspace*{-10mm}
   \subfigure{ \includegraphics[scale=0.23, angle=-90]{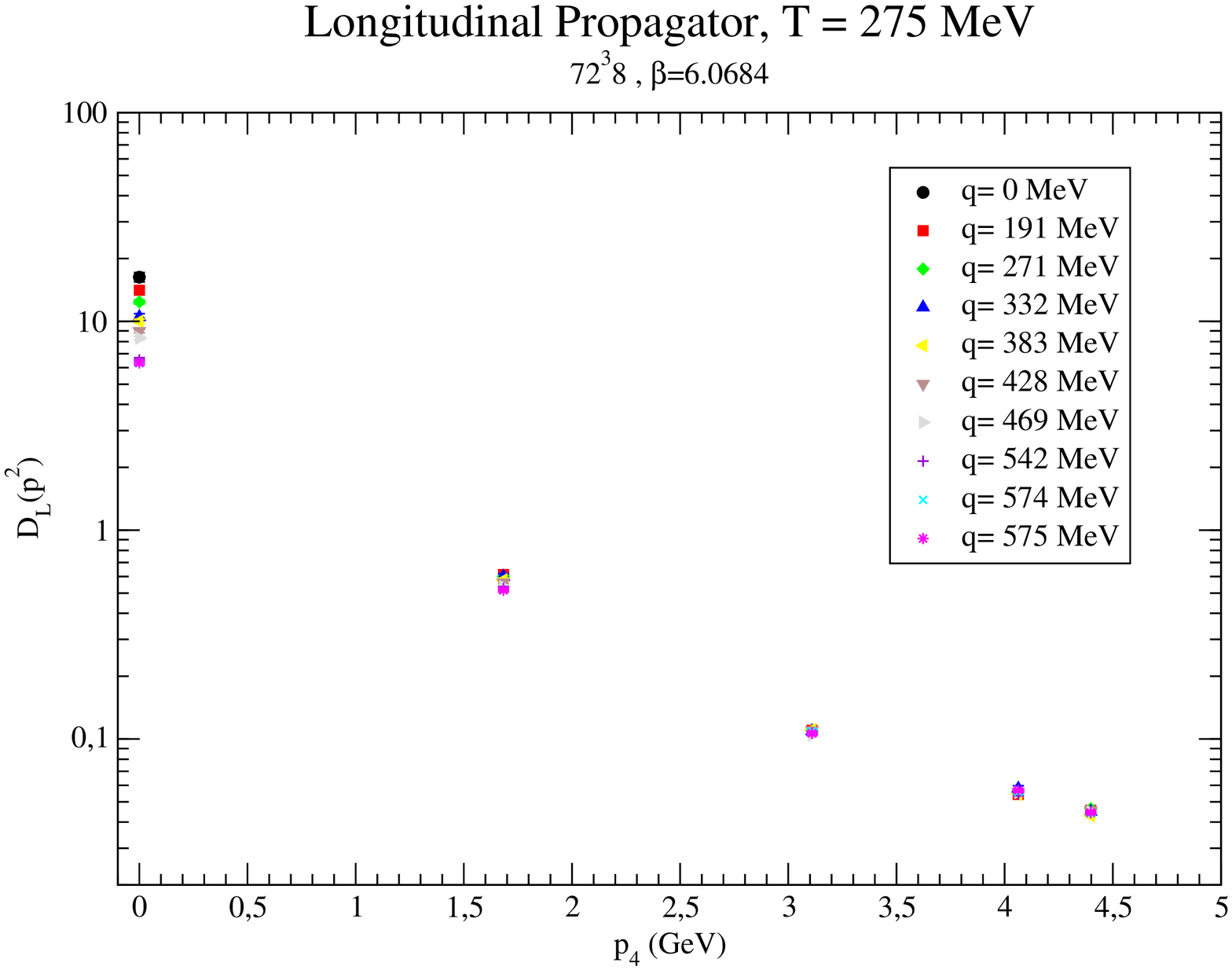} }
  \caption{ $D_{L}(q_4, \vec{q})$ for several spatial momenta $\vec{q}$ and for $T=265$ MeV (left) and $T=275$ MeV (right).}
   \label{q4fig}
\end{figure}

Finally, in Fig. \ref{q4fig} we see how the propagator behaves as a function of $q_4$, for various spatial momenta. 

\section{Gluon spectral densities}

For an Euclidean momentum-space propagator of a (scalar) physical degree of freedom
\begin{displaymath}
\mathcal{G}(p^2)\equiv\braket{\mathcal{O}(p)\mathcal{O}(-p)}
\end{displaymath}
there is a corresponding K\"{a}ll\'{e}n-Lehmann spectral representation 
\begin{displaymath}
\mathcal{G}(p^2)=\int_{0}^{\infty}\d\mu\frac{\rho(\mu)}{p^2+\mu}\,,\qquad \textrm{with }\rho(\mu)\geq0 \textrm{ for } \mu\geq 0\,
\end{displaymath}
where the (positive) spectral density contains information on e.g. the masses of physical states described by the operator $\mathcal{O}$. For physical states, the (positive) spectral density can be determined using e.g. the maximum entropy method (MEM) \cite{mem}. 

However, for the case of unphysical particles like gluons, we need a different method to extract the spectral density, since MEM requires positive spectral densities. In \cite{spectral} we exposed a method based on Tikhonov regularization combined with the Morozov discrepancy principle that allows for positive and negative values for the spectral density. Results were shown for the gluon spectral density at zero temperature. 

Here we are interested to apply the method to the finite temperature case. It is convenient to consider a single spectral function for each spatial momentum:

\begin{equation}
\mathcal{D}(q_4,\vec{q})=\int_{0}^{\infty}\d\mu\frac{\rho(\mu, \vec{q})}{q_4^2+\mu}.
\end{equation}

\begin{figure}[t] 
\centering
\vspace*{0.3cm}
   \includegraphics[scale=0.32]{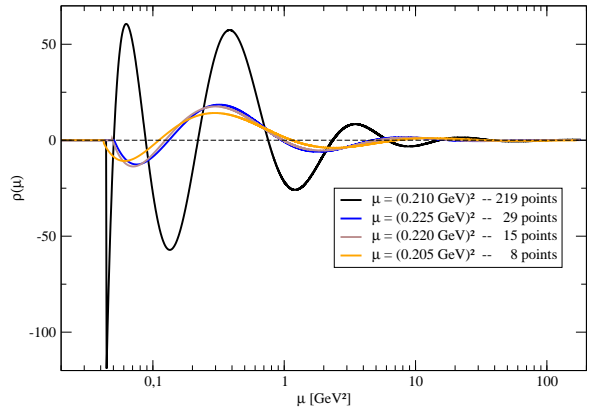} 
  \caption{Gluon spectral densities at $T=0$ for different sets of data points considered in the inversion.}
   \label{spectzero}
\end{figure}

The reader should be aware that for finite $T$, and for the lattice sizes considered, only a small number of Matsubara frequencies are available. In Fig. \ref{spectzero} we consider the inversion of the gluon propagator at zero temperature, considering different numbers of data points.  We conclude that the main features are not affected by the number of data points considered in the inversion. The very preliminary results presented in this work are just for the spatial momentum $\vec{p}=(1,0,0)$.

\begin{figure}[t]
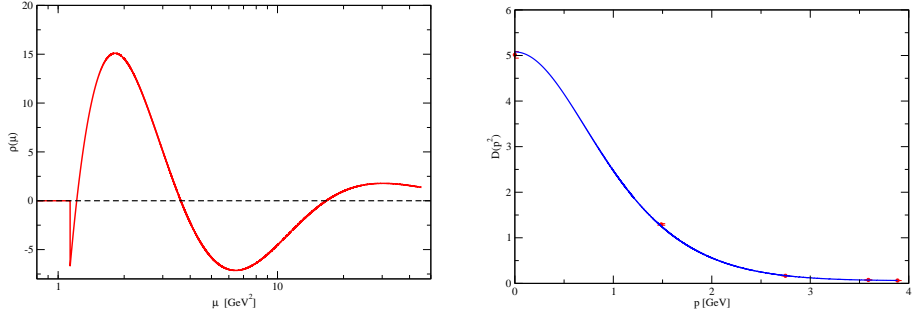
 
   \centering
   \subfigure{ \includegraphics[scale=0.24]{figtemp/rho_trans_T243.eps} } \hspace*{2mm}
   \subfigure{ \vspace{-1.5cm}\includegraphics[scale=0.24]{figtemp/prop_trans_T243.eps} }
  \caption{Spectral density (left) and reconstructed propagator (right) for $D_T$ at $T=243$ MeV.}
   \label{spectral-trans-243}
\end{figure}

\begin{figure}[h] 
   \centering
   \subfigure{ \includegraphics[scale=0.24]{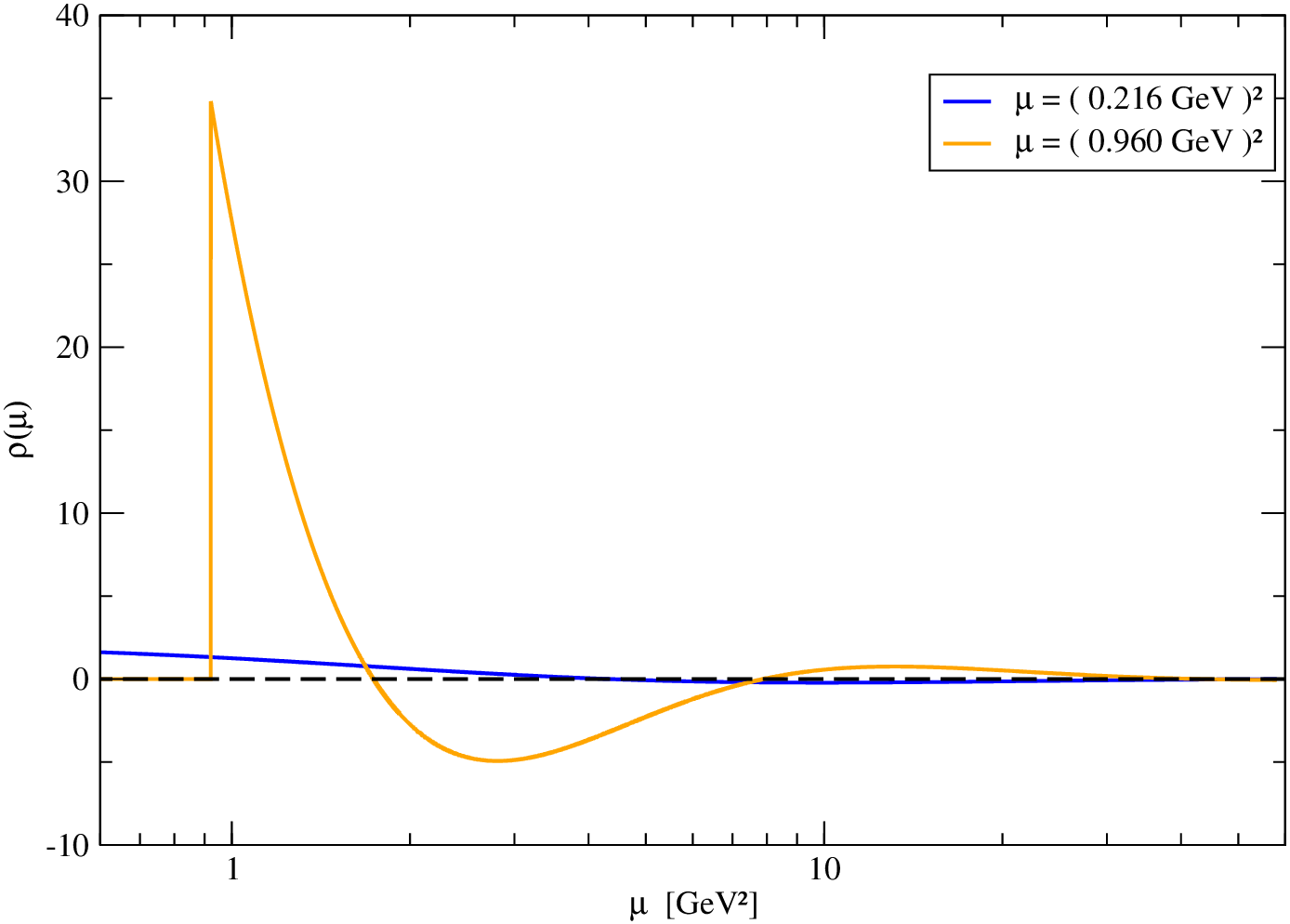} } \,
   \subfigure{ \includegraphics[scale=0.24]{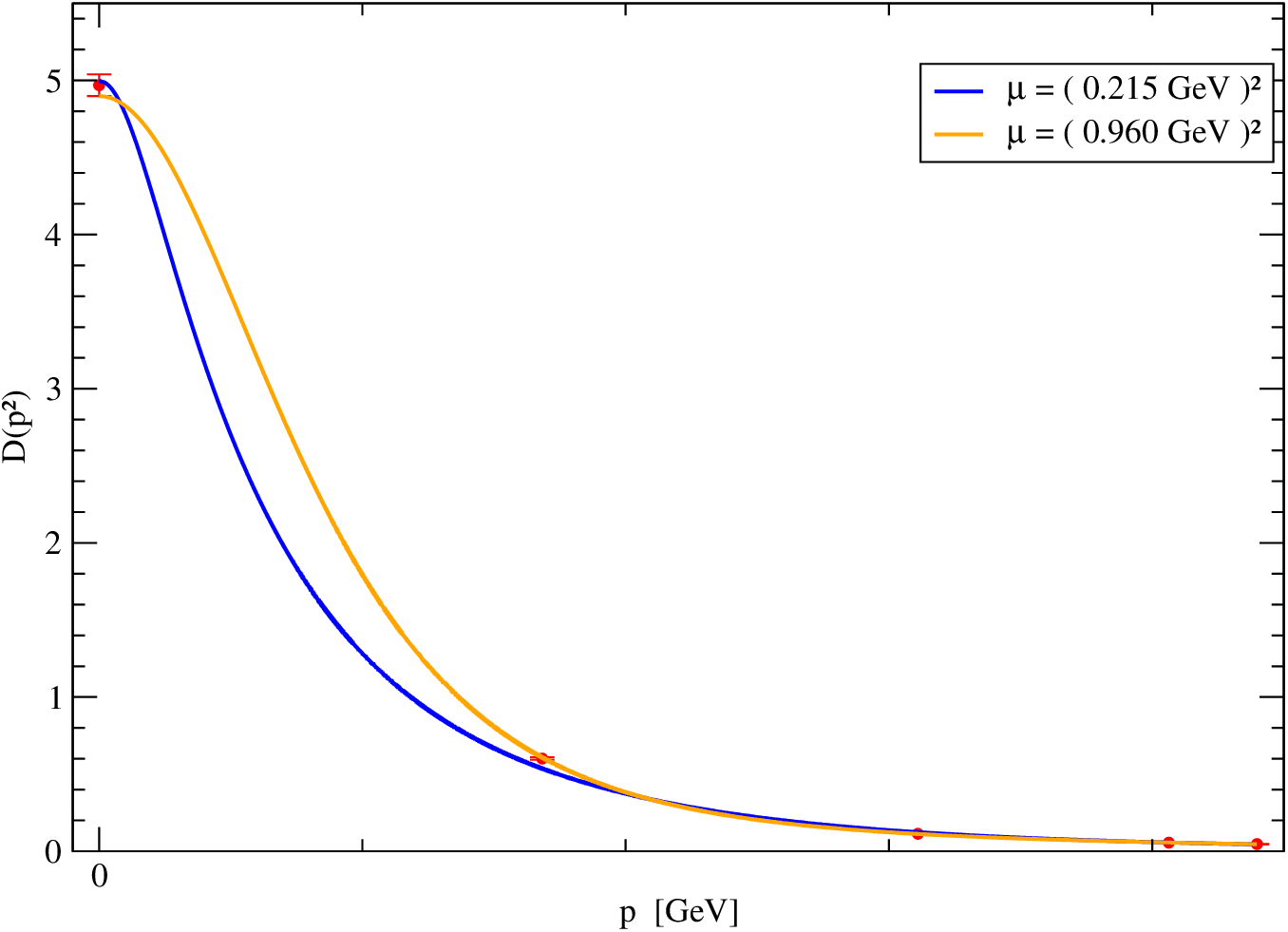} }
  \caption{Spectral density (left) and reconstructed propagator (right) for $D_T$ at $T=275$ MeV.}
   \label{spectral-trans-275}
\end{figure}

The finite $T<T_c$ spectral densities associated to the transverse component 
are similar to the $T=0$ case --- see Fig. \ref{spectral-trans-243}. However, 
above $T_c$ the spectral densities become different, as can be seen in Fig. \ref{spectral-trans-275}. For the longitudinal component  --- see Fig. \ref{spectral-long-260}, the spectral densities have the same pattern for all $T$, different from  the $T=0$ case. 

\begin{figure}[t]
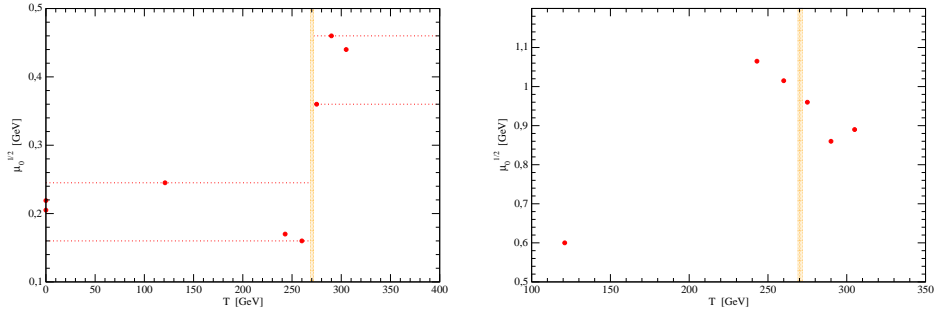
 
   \centering
   \subfigure{ \includegraphics[scale=0.24]{figtemp/IRcutoff_T_long.eps} } \,
   \subfigure{ \includegraphics[scale=0.24]{figtemp/IRcutoff_T_trans.eps} }
  \caption{Infrared cut-offs as functions of the temperature, for $D_L$ (left) and $D_T$ (right).}
   \label{cutoffs}
\end{figure}

In Fig. \ref{cutoffs} we report the dependence of the infrared cut-off $\mu_0$ on the temperature, which seems to be sensitive to the deconfinement phase transition.

Similar results for the spectral density have been obtained recently with an alternative method \cite{iprt}.

\begin{figure}[b] 
   \centering
   \subfigure{ \includegraphics[scale=0.24]{figtemp/rho_long_T260.eps} } \,
   \subfigure{ \includegraphics[scale=0.24]{figtemp/prop_long_T260.eps} }
  \caption{Spectral density (left) and reconstructed propagator (right) for $D_L$ at $T=260$ MeV.}
   \label{spectral-long-260}
\end{figure}

\clearpage

\section*{Acknowledgments}
O. Oliveira, and P. J. Silva acknowledge financial support from FCT Portugal under contract with reference UID/FIS/04564/2016. P. J. Silva acknowledges support by FCT under contracts SFRH/BPD/40998/2007 and SFRH/BPD/109971/2015. The research of M. Roelfs is funded by KU Leuven IF project C14/16/067. The computing time was provided by the Laboratory for Advanced Computing at the University of Coimbra \cite{lca}.

\end{document}